\documentclass[12pt]{article}
\usepackage{latexsym,graphicx}
\usepackage{subfigure}
\usepackage{amssymb}
\usepackage{amsmath}
\usepackage{amscd}
\usepackage{amsthm}
\usepackage[left=2cm,top=2.5cm,right=2.5cm,bottom=1.5cm]{geometry}  
           
\begin{document}

\begin{center}
\large{\bf{Possible Contributions to the Bulk Casimir Energy in Heterotic M-theory.}}
\end{center}

\begin{center}
\normalfont{\bf{Nasr Ahmed$^1$,$^2$}}\\
\normalsize{$^{1}$ Mathematics Department, Faculty of Science, Taibah University, Saudi Arabia.\\$^2$Astronomy Department, National Research Institute of Astronomy and Geophysics, Helwan, Cairo, Egypt\footnote{nkhalifa@taibahu.edu.sa}} 
\end{center}

\begin{abstract}
some possible ways for the study of the contributions of some background fields to the bulk Casimir energy have been probed in the framework of the 5D heterotic M-theory.
\end{abstract}

\section{Introduction}

One of the main theoretical issues in theories with extra dimensions is that of determining their size. If we are interested only in one extra dimension, then the scalar degree of freedom governing the separation is called radion. Some mechanism is always required to fix the size of the extra-dimension and then ensure the stability of the system.\\

A lot of stabilization mechanisms have been proposed in the literature: Introducing a massive scalar field to the bulk \cite{(wise)}, Casimir energy approach \cite{nasr}-\cite{(84)}, Gaugino condensation approach (nonperturbative effects) \cite{nasr2}, \cite{(85)}-\cite{kklt}, flux compactification approach \cite{kklt,Gflux44,Gflux22}. The condition to reach the stabilization is to get the minimum potential.

Instead of introducing an ad-hoc classical interaction between the branes (through the bulk scalar field), Casimir energy of bulk fields may be sufficient to stabilize the radion. Even before branes, Candelas and Weinberg \cite{(84)} found that the quantum effects from matter fields, or gravity, can be used to fix the size of compact extra dimensions. The Casimir potential for untwisted fermions in the warped heterotic $M-$theory background was calculated in \cite{(15)}. The general case of twisted fermions has been done by Ahmed and Moss \cite{nasr}.\\

it has been found that the resultant Casimir energy is not enough to stabilize the radion unless we add other effects such as the (positive) gravitino vacuum energy and gaugino condensates (Ahmed and Moss \cite{nasr2,nasr}). In this paper we try to shed the light on some bulk background fields and their possible contributions to the bulk Casimir Energy.\\

\section{5D heterotic M-theory fields}

In the Horava?itten formulation of M-theory \cite{(5),(5a)}, the gauge fields of the standard model are confined on two 
9-branes located at the end points of an $S^{1}/Z_{2}$ orbifold. The 6 extra dimensions on the branes are compactified on a very small scale, close to the fundamental scale, and their effect on the dynamics is felt through moduli fields, i.e. 5D
scalar fields. A 5D reduction of the Horava?itten theory and the corresponding brane-world
cosmology is given in \cite{(6),(7),(8)}. 

In the 11D theory, the supergravity multiplet consists of the graviton, gravitino and the field $C$. 
The total bulk field content of this 5 dimensional theory is given by the gravity multiplet $(g_{\alpha \beta}, A_{\alpha}, \psi^{i}_{\alpha})$ together with the universal hypermultiplet $(V,\sigma, \zeta, \bar{\zeta})$. $V$ is the Calabi-Yau volume.
After the dualization, the three-form $C_{\alpha \beta \gamma}$ produces a scalar field $\sigma$. The 5 dimensional effective action can be written as \cite{(7)}
\begin{equation}
S_{5}=S_{bulk}+S_{bound}
\end{equation}
Where
\begin{eqnarray}
S_{bulk}&=&\frac{-1}{2\kappa_{5}^{2}}\int_{{\cal M}_{5}}\sqrt{-g} \left[{\cal R}+\frac{3}{2}\bar{F}_{\alpha \beta}\bar{F}^{\alpha \beta}+\frac{1}{\sqrt{2}}\epsilon^{\alpha \beta \gamma \delta \epsilon} A_{\alpha}\bar{F}_{\beta \gamma}\bar{F}_{\delta 
\epsilon} \right. + \\ \nonumber
&& \left. \frac{1}{2V^{2}}\partial_{\alpha}V \partial^{\alpha}V+  
\frac{1}{2V^{2}}\left[(\partial_{\alpha}\sigma - i(\zeta \partial_{\alpha}\bar{\zeta}-\bar{\zeta}\partial_{\alpha}\zeta)-2\alpha \epsilon (x^{11}) A_{\alpha})\right]^2\right.  \\ \nonumber
&& \left. +\frac{2}{V} \partial_{\alpha}\zeta \partial^{\alpha}\bar{\zeta}+\frac{\alpha^{2}}{3V^{2}} \right]
\end{eqnarray}
And
\begin{eqnarray} \label{boundddd} 
S_{bound}&=&\frac{\sqrt{2}}{\kappa_{5}^{2}}\left[\int_{M_{4}^{(1)}}\sqrt{-g}V^{-1}\alpha -\int_{M_{4}^{(2)}}\sqrt{-g}V^{-1}\alpha \right]\\  \nonumber
&-&\frac{1}{16\pi \alpha_{GUT}}\sum_{i=1}^{2}\int_{M_{4}^{(i)}}\sqrt{-g}\left(V \text{tr}F_{\mu\nu}^{(i)}F^{(i)\mu\nu} - \sigma \text{tr}F_{\mu\nu}^{(i)}\widetilde{F}^{(i)\mu\nu}\right).
\end{eqnarray}
where $\widetilde{F}^{(i)\mu\nu}=\frac{1}{2}\epsilon^{\mu \nu \rho \sigma} F_{\rho \sigma}^{(i)}$ and the expansion coeffecients $\alpha_i$ are
\begin{equation}
\alpha_i=\frac{\pi}{\sqrt{2}}\left(\frac{\kappa}{4\pi}\right)^{2/3}\frac{1}{v^{2/3}}\beta_i, \quad \beta_i=-\frac{1}{8\pi^2}\int_{C_i}tr({\cal R}\wedge{\cal R}).
\end{equation}
with the Calabi-Yau volume $V$ defined as
\begin{equation}
V=\frac{1}{v}\int_{X}\sqrt{g^{(6)}}
\end{equation}
where $g^{(6)}$ is the determinant of the Calabi-Yau metric.

When we don't consider the $zeta$-field, the C-field itself is zero. When the $zeta$-field is considered, the C-field is described through the following relations:
\begin{equation}
C_{ABC}=\frac{\sqrt{2}}{24}\epsilon w_{ABC}+\frac{1}{6}\zeta \varepsilon_{ABC} \label{cfield}
\end{equation}
\begin{equation}
C_{\alpha AB}=\frac{1}{18}h B_{[AB~}n_{\alpha]}
\end{equation}
Where $n_{\alpha}$ is a unit vector in the direction of $dz$, $\varepsilon_{ABC}$ is a 3-form where
\begin{equation}
dB=\varepsilon
\end{equation}
on $\partial M$, we have the following boundary conditions for the C-field:
\begin{equation}
C_{ABC}=\frac{\sqrt{2}}{24}\epsilon w_{ABC}-\frac{\sqrt{2}}{48}\epsilon \bar{\chi} \Gamma_{ABC} \chi~~~~on~~ z=0
\end{equation}
\begin{equation}
C_{ABC}=\frac{\sqrt{2}}{24}\epsilon w_{ABC}+\frac{\sqrt{2}}{48}\epsilon \bar{\chi} \Gamma_{ABC} \chi~~~~on ~~z=1
\end{equation}
Where the term $\frac{\sqrt{2}}{48}\epsilon \bar{\chi} \Gamma \chi$ belongs to the gaugino condensates. Comparing the boundary conditions with (\ref{cfield}) we have
\begin{equation}
\epsilon \bar{\chi} \Gamma_{ABC} \chi=\frac{-8}{\sqrt{2}}\zeta \varepsilon_{ABC} ~~~~on~~ z=0
\end{equation}
\begin{equation}
\epsilon \bar{\chi} \Gamma_{ABC} \chi=\frac{8}{\sqrt{2}}\zeta \varepsilon_{ABC} ~~~~on~~ z=1
\end{equation}

\subsection{Background $\zeta$ field and graviphoton} \label{contrib}
Gaugino condensates mean that the gaugino vacuum expectation value is not zero which breaks supersymmetry. \\
The gaugino condinsates lead to a background $\zeta$ field. So, we have a background $V$ field (the size of Calabi-Yau space), a background $\zeta$ field and a background metric. In order to investigate the contribution of the background $\zeta$ field we take the following nonlinear sigma model lagrangian 
\begin{equation}
L=-\frac{1}{4V^{2}}(\partial V)^{2}-\frac{1}{4V^{2}}\left(\partial \sigma-i(\zeta \partial \bar{\zeta}-\bar{\zeta}\partial \zeta)\right)^{2}-\frac{1}{V}(\partial \zeta)(\partial \bar{\zeta})-U  \label{nonlinear}
\end{equation}

Where $U=\frac{\alpha^{2}}{6V^{2}}$.\\

We try to include the gauge vector field $A$ (graviphoton) of the gravity multiplet $(g_{\mu\nu}, \psi_{\mu}, A_{\mu})$ into the universal hypermultiplet and see if it has a little contribution. We do that by making the following transformation for the universal hypermultiplet lagrangian (\ref{nonlinear})
\begin{equation}
\partial \sigma \rightarrow \partial \sigma-2\alpha A
\end{equation}
So, the lagrangian becomes:
\begin{equation}\label{oldl}
L=-\frac{1}{4V^{2}}(\partial V)^{2}-\frac{1}{4V^{2}}((\partial \sigma-2\alpha A)-i(\zeta \partial \bar{\zeta}-\bar{\zeta}\partial \zeta))^{2}-\frac{1}{V}(\partial \zeta))(\partial \bar{\zeta})-U 
\end{equation}
We are interested only in terms containing $A$. That means 
\begin{equation}
L_{A}=-\frac{1}{V^{2}}\left(\alpha^{2}A^{2}-\alpha A (\partial \sigma)+i \alpha A(\zeta \partial \bar{\zeta}-\bar{\zeta}\partial \zeta)\right)
\end{equation}
We make the following perturbations for the universal hypermultiplets:
\begin{eqnarray}
V \rightarrow V+\epsilon \widehat{V}~~~,~~~\zeta \rightarrow \zeta+\epsilon \widehat{\zeta }~~~,~~~ \bar{\zeta}\rightarrow \bar{\zeta}+\epsilon \widehat{\bar{\zeta}} \\  \nonumber
\sigma \rightarrow \epsilon \widehat{\sigma}~~~,~~~ A \rightarrow \epsilon A \\  \nonumber
\end{eqnarray}
where $\epsilon$ is small. After some calculations we find
\begin{equation}
L_{A}=\epsilon^{2}V^{-2}\left(\alpha A (\partial \widehat{\sigma})-\alpha^{2} A^{2}-i \alpha A \zeta \partial(\widehat{\bar{\zeta}}-\widehat{\zeta })\right)
\end{equation}
Integration by parts for the cross terms we get
\begin{eqnarray}
L_{A}&=&\epsilon^{2}\alpha V^{-2}\left(2V^{-1}(\partial_{\mu}V)A^{\mu}\widehat{\sigma}-(\partial_{\mu}A^{\mu})\widehat{\sigma} \right.\\  \nonumber
&&\left.-\alpha A^{2}-2iV^{-1}(\partial_{\mu}V)A^{\mu}\zeta(\widehat{\bar{\zeta}}-\widehat{\zeta })
+(\partial_{\mu}A^{\mu})\zeta(\widehat{\bar{\zeta}}-\widehat{\zeta })+2A^{\mu}(\widehat{\bar{\zeta}}-\widehat{\zeta })\partial_{\mu}\zeta\right)
\end{eqnarray}
The lagrangian (\ref{oldl}) has a global symmetry so it leads to a nonrenormalizable theory. On the contrary, $L_{A}$ is gauge invariant with local $U(1)$ symmetry. We need an expression for the lagrangian in which the $A$ field is separated so we can see its own contribution. Let's introduce a gauge fixing term, which breaks the gauge symmetry, to omit the undesirable cross terms:
\begin{equation}
L_{GF}=-F^{2}\equiv\left(\partial_{\mu}A^{\mu}-\frac{1}{2}\alpha V^{-2}\widehat{\sigma}+2w_{;\mu}A^{\mu}+\frac{1}{2}\alpha
iV^{-2}\zeta(\widehat{\bar{\zeta}}-\widehat{\zeta })\right)^{2}
\end{equation}
Where $A^{\mu}$ is a gauge field and $w$ is a conformal factor defined by
\begin{equation}
\Omega(\sigma)=e^{(1-\sigma)w(z)}
\end{equation}
So that flat space is at $\sigma=0$ and the 5D physical metric

\begin{equation}
ds^2_5=\left(\frac{z}{z_1}\right)^{2/5}\left(g_{\mu\nu}dx^{\mu}dx^{\nu}+dz^2\right)
\end{equation}

is at $\sigma=1$. Now we can write

\begin{eqnarray}
L_{A}+L_{GF}&=&\epsilon^{2}\alpha V^{-2}\left(2V^{-1}(\partial_{\mu}V)A^{\mu}\widehat{\sigma}-\alpha A^{2}-2iV^{-1}(\partial_{\mu}V)A^{\mu}\zeta(\widehat{\bar{\zeta}}-\widehat{\zeta }) \right.\\  \nonumber
&+&\left.2A^{\mu}(\widehat{\bar{\zeta}}-\widehat{\zeta })\partial_{\mu}\zeta\right)
-(\partial_{\mu}A^{\mu})^{2}+\frac{1}{4}\alpha^{2}V^{-4}\widehat{\sigma}^{2}-4(w_{;\mu}A^{\mu})^{2}+\frac{1}{4}V^{-4}\alpha^{2}\zeta^{2}(\widehat{\bar{\zeta}}-\widehat{\zeta })^{2} \\  \nonumber
&-&2iV^{-2}\alpha \zeta(\widehat{\bar{\zeta}}-\widehat{\zeta })w_{;\mu}A^{\mu}-4w_{;\mu}A^{\mu}(\partial_{\alpha}A^{\alpha})-2\alpha V^{-2}\widehat{\sigma}w_{;\mu}A^{\mu}+\frac{i}{2} \alpha^{2}V^{-4}\widehat{\sigma}\zeta(\widehat{\bar{\zeta}}-\widehat{\zeta }).
\end{eqnarray} 

So the gauge fixing term removed only some cross terms but not all of them. A possible reason is that this is due to a wrong choice for the gauge fixing term. The addition for the faddeev-Popov term will not help. We try to perturb the $\zeta$ field in addition to taking the gravitational part
 
\begin{equation}\label{actttt}
\frac{1}{2k^{2}}\int R \sqrt{g} d^{5}x 
\end{equation}

into account. We will not perturb $R$ as we have been assuming a fixed background metric. The coupling between the universal hypermultiplet and gravity multiplet can't be taken into account in (\ref{actttt}). A coupling term  like $CR \zeta^{2}$ is not allowed in the lagrangian as it breaks the supersymmetric structure of the theory, Although it should be taken into account. Being working in the framework of heterotic M theory means we are restricted to start with a supersymmetric theory and then leave the supersymmetry to be broken dynamically. We have no freedom to add other terms like for example a quadratic term in the curvature (the famous gauss Bonnet term). We now have the lagrangian
\begin{eqnarray} \label{lll}
2k^{2}L=R-\frac{1}{2}V^{-2}(\partial_{a} V)(\partial^{b} V)-\frac{1}{3} \alpha^{2} V^{-2}~~~~~~~~~~~~~~~~~~~~~~~~~~~~~~~~~~~~~~~~~~~~~~~~~~~~~~~~\\  \nonumber
-\frac{1}{2}V^{-2}[\partial \sigma-i(\zeta(\partial \bar{\zeta}+\bar{h}n)
-\bar{\zeta}(\partial \zeta +hn))]_{a}[\partial \sigma-i(\zeta(\partial \bar{\zeta}+\bar{h}n)
-\bar{\zeta}(\partial \zeta +hn))]^{b}~~~~~~~\\  \nonumber
-2V^{-1}(\partial \zeta+hn)_{a}(\partial \bar{\zeta}+\bar{h}n)^{b}~~~~~~~~~~~~~~~~~~~~~~~~~~~~~~~~~~~~~~~~~~~~~~~~~~~~~~~~~~~~~~~~~~~~~~~   
\end{eqnarray} 

Where $h$ now is a vector vanish everywhere except in the z-direction.

\subsection{Using the Hamiltonian constraint}

As it is well known, the dynamics is controlled by the Hamiltonian constraint $H=0$ which is the time-time part of the Einstein equations in the 3+1 formalism of General Relativity. We recall that in the 3+1 formalism, Einstein equations are separated into 4 so-called constraint equations corresponding to the time-time and space-time parts, and 6 evolution equations corresponding to the space-space part. The equation obtained from the time-time part of the Einstein equations is called the Hamiltonian constraint, and the ones obtained from the space-time parts are known as the momentum constraints. When written in terms of the extrinsic curvature of the spatial hypersurfaces, the constraint equations have no time derivatives. That means they represent relations that must be satisfied at any given time.\\
Now we take the following background metric ansatz
\begin{equation} \label{mett1}
ds^{2}=e^{-2\omega(z)}(dz^{2}+\eta_{\mu\nu}dx^{\mu}dx^{\nu})
\end{equation}
The gravitational Action now is
\begin{equation}
S_{G}=\frac{1}{2\kappa^{2}}\int e^{-3\sigma}(8\omega^{''}-12\omega^{'2})d^{(5)}x
\end{equation}
Where the primes indicate differentiation with respect to $z$ as all background fields are a function of $z$. After integration by parts, the action could be written as
\begin{equation}
S_{G}=\frac{6}{\kappa^{2}}\int e^{2\omega}\omega^{'2}d^{(5)}V
\end{equation}
Where we have $d^{(5)}V=\sqrt{g^{(5)}}d^{(5)}x$ and $\sqrt{g^{(5)}}=-e^{-5\omega}$. That means  
\begin{equation}
\tilde{R}=12\omega^{'2}e^{2\omega}
\end{equation}
The lagrangian (\ref{lll}) can now be written as:
\begin{eqnarray}
&&L=\frac{-6}{\kappa^{2}}\omega^{'2}e^{-3\omega}+\frac{1}{2}\phi^{'2}e^{-3\omega}+\frac{1}{\kappa^{2}}Ue^{-5\omega}+\frac{1}{4\kappa^{2}}V^{-2}e^{-3\omega}\left(\sigma^{'} \right.\\   \nonumber
&&\left.-i\left(\zeta(\bar{\zeta^{'}}+\bar{h})
-\bar{\zeta}(\zeta^{'} +h)\right)\right)^{2}+\frac{V^{-1}}{\kappa^{2}}e^{-3\omega}(\zeta^{'}+h)(\bar{\zeta^{'}}+\bar{h})
\end{eqnarray}
Where for example $\partial_{a}V \partial^{b}V \equiv \partial_{a}V\partial_{b}Vg^{ab}=e^{2\omega}(\partial V)^{2}$, $V=e^{\sqrt{2}\kappa\phi(z)}$ and $n=1$ in the direction of $z$. The Hamiltonian constraint we are going to use is
\begin{equation}
H\equiv N\frac{\partial L}{\partial N}=0   \label{constra}
\end{equation}
Where $N$ is the lapse function that we are going to take into consideration. We perform a conformal rescaling on the z-direction using the lapse function as a conformal factor.
\begin{equation}
ds^{2}=e^{-2\omega}(\eta_{\mu\nu}dx^{\mu}dx^{\nu}+N^{2}dY^{2})
\end{equation}
Where:
\begin{equation}
NdY=dz
\end{equation}
The unit spacelike normal to the brane surface is given as
\begin{equation}
n_{A}=N \partial_{A}Y
\end{equation}
satisfying the normalization condition
\begin{equation}
g_{AB}n^{A}n^{B}=1
\end{equation}
Where the lapse function is the scalar function
\begin{equation}
N=\left|g_{AB}\partial_{A}Y\partial_{B}Y\right|^{-1/2}
\end{equation}
For the metric determinant we have
\begin{equation}
\sqrt{-g}=Ne^{-5\omega}
\end{equation}
The lagrangian now becomes

\begin{eqnarray}
L&=&\frac{-6}{\kappa^{2}}N^{-1}\omega^{'2}e^{-3\omega}+\frac{1}{2}N^{-1}\phi^{'2}e^{-3\omega}+\frac{1}{\kappa^{2}}NUe^{-5\omega}\\   \nonumber
&+&\frac{1}{4\kappa^{2}}N^{-1}V^{-2}e^{-3\omega}\left(\sigma^{'}-i\left(\zeta(\bar{\zeta^{'}}+\bar{h})
-\bar{\zeta}(\zeta^{'} +h)\right)\right)^{2}\\   \nonumber
&+&\frac{V^{-1}}{\kappa^{2}}N^{-1}e^{-3\omega}(\zeta^{'}+h)(\bar{\zeta^{'}}+\bar{h})
\end{eqnarray}

Where the prime denotes $d/dY$.
The Hamiltonian constraint (\ref{constra}) leads to  
\begin{eqnarray}
H&=&\frac{-6}{\kappa^{2}}N^{-1}\omega^{'2}e^{-3\omega}-\frac{1}{2}N^{-1}\phi^{'2}e^{-3\omega}+\frac{1}{\kappa^{2}}NUe^{-5\omega}\\   \nonumber
&-&\frac{1}{4\kappa^{2}}N^{-1}V^{-2}e^{-3\omega}\left(\sigma^{'}-i\left(\zeta(\bar{\zeta^{'}}+\bar{h})
-\bar{\zeta}(\zeta^{'} +h)\right)\right)^{2}\\   \nonumber
&-&\frac{V^{-1}}{\kappa^{2}}N^{-1}e^{-3\omega}(\zeta^{'}+h)(\bar{\zeta^{'}}+\bar{h})=0~~~~~~~~~~~~~~~~~~
\end{eqnarray}

The momenta of the background fields $p_{i}=\frac{\partial L}{\partial q^{'}_{i}}$are

\begin{eqnarray} \label{mom}
p_{\phi}&=&\phi^{'}N^{-1}e^{-3\omega}\\
p_{\omega}&=&\frac{-12}{\kappa^{2}}\omega^{'}N^{-1}e^{-3\omega}\\
p_{\sigma}&=&\frac{V^{-2}}{2\kappa^{2}}N^{-1}e^{-3\omega}\left(\sigma^{'}-i\left(\zeta(\bar{\zeta^{'}}+\bar{h})
-\bar{\zeta}(\zeta^{'} +h)\right)\right)\\
p_{\zeta}&=&i\bar{\zeta}p_{\sigma}+\frac{V^{-1}}{\kappa^{2}}N^{-1}e^{-3\omega}(\bar{\zeta^{'}}+\bar{h})\\
p_{\bar{\zeta}}&=&-i\zeta p_{\sigma}+\frac{V^{-1}}{\kappa^{2}}N^{-1}e^{-3\omega}(\zeta^{'}+h)
\end{eqnarray}
We express the Hamiltonian in terms of the quadratic momenta as 
\begin{eqnarray}
H&=&-\frac{\kappa^{2}}{24}Ne^{3\omega}p_{\omega}^{2}+\frac{1}{2}Ne^{3\omega}p_{\phi}^{2}-\frac{1}{\kappa^{2}}NUe^{-5\omega}
-\kappa^{2}V^{2}Ne^{3\omega}P_{\sigma}^{2}\\  \nonumber
&-&\kappa^{2}VNe^{3\omega}\left(P_{\zeta}^{2}+i\zeta P_{\zeta}P_{\sigma}-i\bar{\zeta}P_{\bar{\zeta}}P_{\sigma}+\zeta^{2}P_{\sigma}^{2}\right)
\end{eqnarray}
Since $\sigma$ is a cyclic coordinate, this implies that $p_{\sigma}$ is constant. We argue also that the quantity $(P_{\zeta}-i\bar{\zeta}P_{\sigma})$ is constant.
Now we would like to know the form of the Hamiltonian on the boundary. That means We need to know the boundary conditions for $\phi, P_{\omega}$ and $P_{\phi}$. For the conformally flat background metric and for the flat brane case, $\partial z_{1}=0$ so we have:
\begin{equation}
n=e^{\omega}\partial_{z}
\end{equation}
The boundary conditions for $\phi$ and $\omega$ could be drived by varying the following actions with respect to $\phi$ and $\omega$ respectively,
\begin{eqnarray}
S&=&S_{bulk}+S_{boundary}\\  \nonumber
&=&\int_{M}L_{bulk} \sqrt{-g^{(5)}}+\int_{\partial M}L_{boundary} \sqrt{-g^{(4)}}
\end{eqnarray}
Where
\begin{eqnarray}
L_{bulk}&=& \frac{1}{2 \kappa^{2}}\left(R-\frac{1}{2}V^{-2}\partial_{\alpha}V \partial^{\alpha}V-\frac{1}{3}V^{-2}\right)  , ~~~~~~~~~V=e^{\sqrt{2}\kappa \phi}\\
L_{boundary}&=& \frac{1}{\kappa^{2}}(K\pm U) , ~~~~~~~~~~~~~~~~~~~~~~~~~~~~~~~~~~~~~~~~~U=\frac{\alpha}{\sqrt{2}}V^{-1}~~~~~\label{ord}
\end{eqnarray}
This yields for $\phi$,
\begin{equation}
n.\nabla \phi =\frac{\alpha}{\kappa}V^{-1}
\end{equation}   
Which implies 
\begin{equation}
\phi^{'}=\frac{\alpha}{\kappa}e^{-\omega}V^{-1}N
\end{equation}
Using (\ref{mom}), the momentum for the $\phi-field$ on the boundary is
\begin{equation}
p_{\phi}=\frac{\alpha}{\kappa}e^{-4\omega}V^{-1}
\end{equation}
For K, we have for the case of flat brane
\begin{equation}
K=4\omega^{'}e^{\omega}~~~~~~~~~~~~~~~~~~at~ z=z_{1} 
\end{equation}
\begin{equation}
K=-4\omega^{'}e^{\omega}~~~~~~~~~~~~~~~at~ z=z_{2} 
\end{equation}
Now we use (see \cite{(8)})
\begin{equation}
K=4\omega^{'}e^{\omega}=\frac{-2\sqrt{2}}{3}\alpha V^{-1}
\end{equation}
So
\begin{equation}
\omega^{'}=-\frac{\alpha}{3\sqrt{2}}e^{-\omega}V^{-1}
\end{equation}
The momentum for the $\omega-field$ on the boundary could be written as
\begin{equation}
p_{\omega}=\frac{2\sqrt{2}}{\kappa^{2}}\alpha V^{-1}e^{-4\omega}
\end{equation}
Then, the Hamiltonian on the boundary is

\begin{eqnarray}
H&=&\frac{-\alpha^{2}}{3\kappa^{2}}V^{-2}Ne^{-5\omega}+\frac{\alpha^{2}}{2\kappa^{2}}V^{-2}Ne^{-5\omega}-\frac{\alpha^{2}}{6\kappa^{2}}V^{-2}Ne^{-5\omega}\\ \nonumber
&-&\kappa^{2}VNe^{3\omega}(VP_{\sigma}^{2}+P_{\zeta}^{2}+iP_{\sigma}(\zeta P_{\zeta}-\bar{\zeta}p_{\bar{\zeta}})+\zeta^{2}P_{\sigma}^{2})
\end{eqnarray}

The first three terms cancel. We then have

\begin{equation}
\kappa^{2}VNe^{3\omega}\left(VP_{\sigma}^{2}+(P_{\zeta}-i\bar{\zeta}P_{\sigma})(P_{\bar{\zeta}}+i\zeta P_{\sigma})\right)=0
\end{equation}

The quantity $\left(VP_{\sigma}^{2}+(P_{\zeta}-i\bar{\zeta}P_{\sigma})(P_{\bar{\zeta}}+i\zeta P_{\sigma})\right)$ is a positive definite quantity, this implies that $P_{\sigma}=0$. Since $P_{\sigma}$ is constant, it vanishes everywhere. Similarly $p_{\zeta}=0$.
The next step is to solve for $\omega(z)$ and $\phi(z)$.
The Hamiltonian now is
\begin{equation}
H=-\frac{\kappa^{2}}{24}Ne^{3\omega}p_{\omega}^{2}+\frac{1}{2}Ne^{3\omega}p_{\phi}^{2}-\frac{\alpha^{2}}{6\kappa^{2}}Ne^{-5\omega-2\sqrt{2}\kappa \phi}
\end{equation}
We can choose $N$ freely.
If we choose $N=e^{-3\omega}$ we get
\begin{equation}
H=-\frac{\kappa^{2}}{24}p_{\omega}^{2}+\frac{1}{2}p_{\phi}^{2}-\frac{\alpha^{2}}{6\kappa^{2}}e^{-8\omega-2\sqrt{2}\kappa \phi}
\end{equation}
Which is a Cyclic Hamiltonian as it doesn't depend on the coordinates $\phi$ or $\omega$ independently, but on a combination of them i.e.
\begin{equation}
H\equiv H(P_{\omega}, P_{\phi},-8\omega-2\sqrt{2}\kappa \phi )
\end{equation}
Using the Hamiltonian equations, we get
\begin{equation}
P^{'}_{\phi}=-\frac{\partial H}{\partial \phi}
=-(\frac{\partial H}{\partial \bar{V}}\frac{\partial \bar{V}}{\partial \phi})=\frac{\sqrt{2}\alpha^{2}}{3\kappa}\bar{V}~
\end{equation}
\begin{equation}
P^{'}_{\omega}=-\frac{\partial H}{\partial \omega}
=-(\frac{\partial H}{\partial \bar{V}}\frac{\partial \bar{V}}{\partial \omega})=\frac{4\alpha^{2}}{3\kappa^{2}}\bar{V}~~~
\end{equation}
Where by $\bar{V}$ we mean $e^{-8\omega-2\sqrt{2}\kappa \phi}$.
That leads to
\begin{equation}
8P^{'}_{\phi}-2\sqrt{2}\kappa P^{'}_{\omega}=0
\end{equation}
Which implies
\begin{equation}
8P_{\phi}-2\sqrt{2}\kappa P_{\omega}=C
\end{equation}
using the boundary values of $P_{\phi}$ and $P_{\omega}$ the constant=0. So,
\begin{equation} \label{soo}
8P_{\phi}-2\sqrt{2}\kappa P_{\omega}=0  
\end{equation}
Substituting back in $H$, we get
\begin{equation}
3\left(\frac{d\omega}{dz}\right)^{2}e^{8\omega}=\frac{\alpha^{2}}{6}e^{-2\sqrt{2}\kappa \phi}  \label{back}
\end{equation}
Equation (\ref{soo}) leads to
\begin{equation}
\frac{d\phi}{dY}+\frac{3\sqrt{2}}{\kappa}\frac{d\omega}{dY}=0
\end{equation}
The solution is
\begin{equation}
\phi=\phi_{o}-\frac{3\sqrt{2}}{\kappa}\omega
\end{equation}
Where $\phi=\phi_{o}$ at $\omega =0$.
Substituting in (\ref{back}) we get the following differential equation,
\begin{equation}
\frac{d\omega}{dY}e^{-2\omega}=\frac{\alpha}{3\sqrt{2}}e^{-\sqrt{2}\kappa \phi_{o}}
\end{equation}
This gives
\begin{equation}
e^{-2\omega}=\left(1-\frac{\alpha \sqrt{2}}{3} \frac{Y}{V_{o}}\right)
\end{equation}
Making use of $e^{-3\omega}dY=dz$, we finally get
\begin{equation}
e^{-2\omega}=\left(\frac{z}{z_{1}}\right)^{2/5}
\end{equation}
Which is the scaling factor in ($\ref{mett1}$). Similarly we get for $\omega(z)$ and $\phi(z)$:
\begin{equation}
\omega(z)=-\frac{1}{5}\ln(\frac{z}{z_{1}})
\end{equation}
\begin{equation}
\phi(z)=\frac{3\sqrt{2}}{5\kappa}\ln(\frac{z}{z_{1}})+\phi_{o}
\end{equation}

That agrees with solutions in \cite{(15)} for the bulk dilaton and radion fields
\section{Changing the boundary potential}
In M theory, there's a fine tuning between the bulk and boundary potentials to get the flat potential of the radion. If the boundary potential is strong enough then it could be separated into two parts, the first one for the flatness of the radion potential and the second part for the radion stabilzation after summing up with the Casimir potential. Since supersymmetry is not observed in nature, we should incorporate supersymmetry breaking.\\
The idea of including a potential on the positive tension brane coming from SUSY breaking effects has been discussed in [98,99,100]. In [98], the simplest potential is obtained by detuning the boundary potential $U_{B}$ defined by  
\begin{equation}
U_{B}=4k e^{\alpha\psi}
\end{equation}
so that the potential becomes
\begin{equation}
V=\frac{6(T-1)k}{\kappa^{2}_{5}}e^{\alpha\psi}
\end{equation}
Where $T\neq 1$ is a SUSY breaking parameter and $\psi$ is the bulk scalar field. We start by modifying the boundary potential in ($\ref{ord}$). The potential now is multiplied by some constant $C$ The boundary Lagrangian becomes
\begin{equation}
L_{bound}= \frac{1}{\kappa^{2}}(K\pm cU)   ~~~~~~~~~~~~~~~~~~,~~~~~~~~~~~~~~~~~~~U=\frac{\alpha}{\sqrt{2}}V^{-1}~~~~~
\end{equation}
Consequently, the boundary conditions for $\phi$ and $\omega$ will be changed as
\begin{eqnarray}
\phi^{'}&=& \frac{\alpha}{\kappa} ce^{-\omega}V^{-1}N\\
\omega^{'}&=& -\frac{\alpha}{3\sqrt{2}} ce^{-\omega}V^{-1}N
\end{eqnarray}
The momenta on the boundary will be modified to 
\begin{eqnarray}
p_{\phi}&=&\frac{\alpha}{\kappa}ce^{-4\omega}V^{-1}\\
p_{\omega}&=&-\frac{\alpha}{3\sqrt{2}}ce^{-\omega}V^{-1}
\end{eqnarray}
Now the Hamiltonian constraint on the boundary gives
\begin{equation}
\frac{\alpha^{2}(c^{2}-1)N}{6\kappa^{2}}V^{-2}e^{-5\omega}-\kappa^{2}VNe^{3\omega}\left(VP_{\sigma}^{2}+(P_{\zeta}-i\bar{\zeta}P_{\sigma})(P_{\bar{\zeta}}+i\zeta P_{\sigma})\right)=0 \label{tett}
\end{equation}
The above picture suggests the extra term to be connected with the SUSY breaking where the constant $c^{2}$ is a SUSY breaking parameter. SUSY is not broken for $c^{2}=1$. The additional proposed part of the potential then is coming from SUSY breaking. \\
Now we would like to solve for $\phi(z)$, $\omega(z)$, $\sigma(z)$ and $\zeta(z)$. \\
Taking $\zeta$ real and $N=e^{-3w}$ the bulk Hamiltonian could be expressed as
\begin{equation}
H=-\frac{\kappa^{2}}{24}p_{\omega}^{2}+\frac{1}{2}p_{\phi}^{2}-\frac{\alpha^{2}}{6\kappa^{2}}e^{-8\omega-2\sqrt{2}\kappa \phi}-\kappa^{2}V^{2}(\phi)P_{\sigma}^{2}-\kappa^{2}V(\phi)(P_{\zeta}^{2}+\zeta^{2}P_{\sigma}^{2})
\end{equation}
The Hamiltonian equations $q_{i}^{'}=\frac{\partial H}{\partial p_{i}}$ lead to
\begin{eqnarray}
\phi^{'}&=&P_{\phi}\\
w^{'}&=&\frac{-\kappa^{2}}{12}P_{w}\\
\sigma^{'}&=&-2\kappa^{2}P_{\sigma}(V^{2}+V\zeta^{2})\\
\zeta^{'}&=&-2k^{2}VP_{\zeta} \label{zzz}
\end{eqnarray}

And $P_{i}^{'}=-\frac{\partial H}{\partial q_{i}}$ lead to

\begin{eqnarray}
P_{\phi}^{'}&=&-\frac{\alpha^{2}\sqrt{2}}{3\kappa}e^{-8\omega-2\sqrt{2}\kappa \phi}+\sqrt{2}\kappa^{3}P_{\zeta}^{2}e^{\sqrt{2}\kappa \phi}\\
P_{w}^{'}&=&-\frac{4\alpha^{2}}{3\kappa^{2}}e^{-8\omega-2\sqrt{2}\kappa \phi}\\
P_{\sigma}^{'}&=&0\\
P_{\zeta}^{'}&=&2\zeta \kappa^{2} V P_{\sigma}^{2}
\end{eqnarray}

Equations ($\ref{tett}$) Gives
\begin{equation} \label{372}
P_{\zeta}^{2}|_{bound}=\frac{\alpha^{2}(c^{2}-1)e^{-8w}V^{-3}}{6k^{4}}-P_{\sigma}^{2}(V+\zeta^{2})
\end{equation}
That means $P_{\sigma}$ is constant, as a special case we consider this constant equals zero. In this case  the background field $\sigma$ is constant and equation ($\ref{372}$) becomes
\begin{equation}
P_{\zeta}^{2}|_{bound}=\frac{\alpha^{2}(c^{2}-1)e^{-8w}V^{-3}}{6k^{4}}
\end{equation}
Assuming $P_{\zeta}|_{bulk}$ is constant. for $c=1$, $P_{\zeta}|_{bound}$ vanishes. That means $P_{\zeta}$ vanishes everywhere. SUSY breaking generates a varying $P_{\zeta}$.
It follows from equation ($\ref{zzz}$) that
\begin{equation}
\zeta=\frac{2\sqrt{2}\alpha k^{2}P_{\zeta}}{75 z_{1}}(1-z^{4})
\end{equation}
which means that the $\zeta$ field vanishes everywhere for the $c=1$ case.
For $w(z)$ and $\phi(z)$,the following second order differential equations system valid in the bulk

\begin{eqnarray}
\frac{d^{2}\phi}{dz^{2}}&=&-\frac{\alpha^{2}\sqrt{2}}{3\kappa}e^{-8\omega-2\sqrt{2}\kappa \phi}+\sqrt{2}\kappa^{3}P_{\zeta}^{2}e^{\sqrt{2}\kappa \phi}\label{369} \\
\frac{d^{2}w}{dz^{2}}&=&\frac{\alpha^{2}}{9}e^{-8\omega-2\sqrt{2}\kappa \phi}  \label{370}
\end{eqnarray}

We are looking for a solution for the above system. If we set the first brane at $z_{1}$ and the second brane at $z_{2}$, then on the boundary we have: 

\begin{eqnarray}
\phi^{'}(z_{1})&=&\frac{\alpha}{\kappa} ce^{-4\omega(z_{1})-\sqrt{2}\kappa \phi(z_{1})}\\
\phi^{'}(z_{2})&=&\frac{\alpha}{\kappa} ce^{-4\omega(z_{2})-\sqrt{2}\kappa \phi(z_{2})}\\
\omega^{'}(z_{1})&=&-\frac{\alpha}{3\sqrt{2}} ce^{-4\omega(z_{1})-\sqrt{2}\kappa \phi(z_{1})}\\
\omega^{'}(z_{2})&=&-\frac{\alpha}{3\sqrt{2}} ce^{-4\omega(z_{2})-\sqrt{2}\kappa \phi(z_{2})}.
\end{eqnarray}

At first, we consider the case for non-broken SUSY. For this case $c=1$ and $P_{\zeta}=0$, then

\begin{eqnarray}
\frac{d^{2}\phi}{dz^{2}}&=&-\frac{\alpha^{2}\sqrt{2}}{3\kappa}e^{-8\omega-2\sqrt{2}\kappa \phi}\\
\frac{d^{2}w}{dz^{2}}&=&\frac{\alpha^{2}}{9}e^{-8\omega-2\sqrt{2}\kappa\phi}
\end{eqnarray}

The solution satisfiying the boundary conditions and the above equations is:

\begin{eqnarray}
w(z)&=&-\frac{1}{2}\ln(\frac{2z}{5z_{1}})\\
\phi(z)&=&\frac{3}{\sqrt{2}k}\ln(\frac{2z}{5z_{1}})-\frac{1}{\sqrt{2}k}\ln(\frac{z_{o}}{z_{1}})
\end{eqnarray}

\begin{table}
\begin{center}
\begin{tabular}{|l||l|l||l|l|}
\hline
 &\multicolumn{2}{l|}{~~~~~~~~~~~~~~$w(z)$}&\multicolumn{2}{l|}{~~~~~~~~~~~~~~~$\phi(z)$}\\
\cline{2-5}
 &Numerical&Exact&Numerical&Exact\\
\hline\hline
1  &0.458145365&0.458145366&-0.344778990&-0.344778990\\
1.1&0.410490132&0.410490276&-0.243687359&-0.243687279\\
1.2&0.366984621&0.366984587&-0.151397852&-0.151397777\\
1.3&0.326963264&0.326963233&-0.066499732&-0.066499474\\
1.4&0.289909282&0.289909247&~0.012103634&~0.0121037095\\
1.5&0.255412854&0.255412811&~0.085281671&~0.0852817001\\
1.6&0.223143609&0.223143551&~0.153735014&~0.1537351395\\
1.7&0.192831307&0.192831240&~0.218037116&~0.2180372611\\
1.8&0.164252103&0.164252033&~0.278662763&~0.2786629141\\
1.9&0.137218500&0.137218422&~0.336009695&~0.3360098622\\
2  &0.111571862&0.111571775&~0.390414431&~0.3904146167\\
\hline
\end{tabular}
\caption{The numerical and exact values of $w(z)$ and $\phi(z)$ for different points between the first brane at $z_{1}$ and second brane at $z_{2}$.}
\label{55}
\end{center}
\end{table}
The numerical investigation of the system for the general SUSY breaking case $c\neq 1$ shows no solution satisfies the boundary conditions and the field equations. This could be done by plotting the following two quantities
\begin{equation}
B_{\phi}=\phi^{'}(z)-\frac{\alpha}{\kappa} ce^{-4\omega(z)-\sqrt{2}\kappa \phi(z)}
\end{equation}
\begin{equation}
B_{w}=\omega^{'}(z)+\frac{\alpha}{3\sqrt{2}} ce^{-4\omega(z)-\sqrt{2}\kappa \phi(z)}
\end{equation}
We find $B_{w}$ is decreasing forever and $B_{\phi}$ is increasing. Therefore, $B_{w}\neq 0$ and $B_{\phi}\neq 0$ $\forall z\neq z_{1}$. That means the boundary conditions are not satisfied $\forall (z_{1}+\epsilon)$ , $\epsilon >0$. For $c^{2}<1$, $B_{\phi}$ is decreasing. In other words, $B_{\phi}$ doesn't change its sign in the closed interval $[1,2]$ and then it follows from the intermediate value theorem that $B_{\phi}$ has no roots in this interval.\\ 
The same could be shown for $B_{w}$ Where it is decreasing forever for $c^{2}>1$ and increasing for $c^{2}<1$. Again the intermediate value theorem implies that there's no roots.\\
For an analytical investigation, we have
\begin{eqnarray}
B_{\phi}^{'}&=&\phi^{''}-\frac{\alpha c}{\kappa}\frac{\partial}{\partial z}
e^{-4\omega(z)-\sqrt{2}\kappa \phi(z)} \\ \nonumber
&=&\frac{\sqrt{2}\alpha^{2}(c^{2}-1)}{2\kappa}e^{-8\omega(z)-2\sqrt{2}\kappa \phi(z)}>0
\end{eqnarray}
And 
\begin{eqnarray} \label{bbb}
B_{w}^{'}&=& w^{''}+\frac{\alpha c}{3 \sqrt{2}}\frac{\partial}{\partial z}
e^{-4\omega(z)-\sqrt{2}\kappa \phi(z)} \\ \nonumber
&=&\frac{-\alpha^{2}(c^{2}-1)}{9}e^{-8\omega(z)-2\sqrt{2}\kappa \phi(z)}<0
\end{eqnarray}

suppose that 

\begin{eqnarray}
&&(1)~~B_{w}(z_{1})=B_{\phi}(z_{1})=0 ~~~\text{and}~~~  B_{w}(z_{2})=B_{\phi}(z_{2})=0.\\
&&(2)~~\nexists z_{3}~\text{such that}~z_{1}<z_{3}<z_{2}~~~\text{and}~~~B(z_{3})=0.
\end{eqnarray}

Using ($\ref{bbb}$) for $c^{2}>1$, if
\begin{equation}
B^{'}_{w}(z_{1})<0 ~\text{and}~B^{'}_{w}(z_{2})<0 
\end{equation}
That means 
\begin{equation}
\exists \widehat{z}_{1}~ \text{such that} ~B_{w}(\widehat{z}_{1})>0,~\widehat{z}_{1}>z_{1}
\end{equation}
And 
\begin{equation}
\exists \widehat{z}_{2}~ \text{such that} ~B_{w}(\widehat{z}_{2})<0,~\widehat{z}_{2}<z_{2}
\end{equation}

But the intermediate value theorem implies that 

\begin{equation}
\exists z_{3}~\text{such that}~B_{w}(z_{3})=0,~z_{3}\in [\widehat{z}_{1},\widehat{z}_{2}]
\end{equation} 
Which contradicts the basic assumption meaning that there are no roots. For $c^{2}<1$, suppose that
\begin{equation}
B^{'}_{w}(z_{1})>0 ~\text{and}~B^{'}_{w}(z_{2})>0 
\end{equation}
That means 
\begin{equation}
\exists \widehat{z}_{1}~ \text{such that} ~B_{w}(\widehat{z}_{1})<0,~\widehat{z}_{1}>z_{1}
\end{equation}
And 
\begin{equation}
\exists \widehat{z}_{2}~ \text{such that} ~B_{w}(\widehat{z}_{2})>0,~\widehat{z}_{2}<z_{2}
\end{equation}
But the intermediate value theorem implies that 
\begin{equation}
\exists z_{3}~ \text{such that}~B_{w}(z_{3})=0,~z_{3}\in [\widehat{z}_{1},\widehat{z}_{2}]
\end{equation} 
Which is a contradiction with the basic assumption. The same argument could be applied for $B_{\phi}$.
Now we need to solve numerically for the $\phi(z)$, $w(z)$, and $\zeta(z)$ fields for $c \neq 0$ case. The Numerical results are shown in  table (\ref{66}). 
\begin{table}
\begin{center}
\begin{tabular}{|l||l|l||l|l||c|}
\hline &\multicolumn{2}{l|}{~~~~~~~~~~~~$w(z)$}&\multicolumn{2}{l|}{~~~~~~~~~~~$\phi(z)$}&$\zeta$\\
\cline{2-5}
 &$c^{2}=1.1$&$c^{2}=0.9$&$c^{2}=1.1$&$c^{2}=0.9$&\\
\hline\hline
1  &~0&~0&1&1&0\\
1.1&-0.0076142167&-0.0062209919&1.0161521928&1.0131967168&0\\
1.2&-0.0151342925&-0.0123473197&1.0321046825&1.0261926205&0\\
1.3&-0.0225630162&-0.0183812740&1.0478633854&1.3899257049&0\\
1.4&-0.2990306537&-0.0243250594&1.0634339809&1.0516012433&0\\
1.5&-0.0371570068&-0.0301807956&1.0788219145&1.0640231358&0\\
1.6&-0.0443272999&-0.0359505192&1.0940324031&1.0762625678&0\\
1.7&-0.0514163099&-0.0416361924&1.1090704643&1.0883237020&0\\
1.8&-0.0584263102&-0.0472397656&1.1239409206&1.1002105486&0\\
1.9&-0.0653594876&-0.0527628814&1.1386484108&1.1119269739&0\\
2&-0.0722179429&-0.0582074752&1.1531973916&1.1234767015&0\\
\hline
\end{tabular}
\caption{The numerical investigation for $w(z)$ and $\phi(z)$ and $\zeta(z)$ for $c\neq0$. The zero initial value $\zeta(1)=0$ leads to a zero $\zeta(z)$ field in the bulk.}
\label{66}
\end{center}
\end{table}

\section*{Conclusion}

some possible ways for the study of the contributions of some background fields to the bulk Casimir energy have been probed in the framework of the 5D heterotic M-theory. The inclusion of the gauge vector field $A$, the graviphoton, into the universal hypermultiplet has been studied. Using the Hamiltonian constraint, we expressed the Hamiltonian in terms of the quadratic momenta and and we found that $P_{\sigma}$ is constant. We then studied the boundary conditions for $\phi$ $P_{w}$ $P_{\phi}$. We investigated the changing of the boundary conditions through detuning the boundary potential $U_B$. The Hamiltonian constraint leads to an extra term related to susy breaking. The numerical investigation of the system for the general SUSY breaking case shows no solution satisfies the boundary conditions and the field equations.  
 
\section*{Acknowledgment}

The Auther is so grateful to Prof. Ian Moss from the school of Matematics and statistics at Newcastle university for his great help and very useful discussions.

\end{document}